\documentclass{aastex63}

\begin{document}

\title{Discovery of a Universal Correlation For Long and Short GRBs and Its Application on the Study of Luminosity Function and Formation Rate}

\correspondingauthor{Da-Ming Wei}
\email{dmwei@pmo.ac.cn}

\author{Qi Guo}
\affil{Key Laboratory of dark Matter and Space Astronomy, Purple Mountain Observatory, Chinese Academy of Science \\
	 Nanjing, Jiangsu 210008, China.}
\affil{School of Astronomy and Space Science, University of Science and Technology of China \\ 
	Hefei, Anhui 230026, China.}

\author{Da-Ming Wei}
\affil{Key Laboratory of dark Matter and Space Astronomy, Purple Mountain Observatory, Chinese Academy of Science \\
	Nanjing, Jiangsu 210008, China.}
\affil{School of Astronomy and Space Science, University of Science and Technology of China \\ 
	Hefei, Anhui 230026, China.}

\author{Yuan-Zhu Wang}
\affil{Key Laboratory of dark Matter and Space Astronomy, Purple Mountain Observatory, Chinese Academy of Science \\
	Nanjing, Jiangsu 210008, China.}

\author{Zhi-Ping Jin}
\affil{Key Laboratory of dark Matter and Space Astronomy, Purple Mountain Observatory, Chinese Academy of Science \\
	Nanjing, Jiangsu 210008, China.}

\begin{abstract}

Gamma-ray bursts (GRBs) are known to be the most violent explosions in the universe, and a variety of correlations between observable GRB properties have been proposed in literature, but none of these correlations is valid for both long GRBs and short GRBs. In this paper we report the discovery of a universal correlation which is suitable for both long and short GRBs using three prompt emission properties of GRBs, i.e. the isotropic peak luminosity $L_{\rm iso}$, the peak energy of the time-integtated prompt emission spectrum $E_{\rm peak}$, and the "high signal" timescale $T_{\rm 0.45}$, $L_{\rm iso} \propto E_{\rm peak}^{1.94} T_{0.45}^{0.37}$. This universal correlation just involves properties of GRB prompt emission and does not require any information of afterglow phase, which can be used as a relatively unbiased redshift estimator.  Here we use this correlation to estimate the pseudo–redshifts for short Gamma Ray Bursts and then use Lynden-Bell  method to obtain a non-parametric estimate of their luminosity function and formation rate. The luminosity function is $\psi(L_0)\propto{L_0^{-0.63\pm{0.07}}}$ for dim SGRBs and $\psi(L_0)\propto{L_0^{-1.96\pm{0.28}}}$ for bright SGRBs, with the break point $6.95_{-0.76}^{+0.84}\times10^{50} erg/s$. The local formation rate of SGRBs is about 15 events $\rm Gpc^{-3}yr^{-1}$ . This universal correlation may have important implications for GRB physics, implying that the long and short GRBs should share similar radiation processes.

\end{abstract}


\section{Introduction}

Gamma-ray Bursts (GRBs) are short, intense gamma-ray flashes that are by far the most violent explosions in the universe \citep{2006RPPh...69.2259M,2007ChJAA...7....1Z,2009ARA&A..47..567G,2015PhR...561....1K}. A lot of researches have shown that GRBs can be divided into two categories, the long Gamma-ray bursts (LGRB) with duration $\rm T_{90} > 2s$ and the short Gamma-ray bursts (SGRB) with $\rm T_{90} \leq 2s$ \citep{1993ApJ...413L.101K}. For different kinds of GRBs, their origins are also different. The progenitors of LGRBs are regarded as collapsed massive stars \citep{1993ApJ...405..273W}, while SGRBs are related to the coalescence of compact objects such as binary neutron stars or neutron star-black holes \citep{2007PhR...442..166N,2017PhRvL.119p1101A,2017ApJ...848L..14G,2017ApJ...848..115H}.

A variety of correlations between observable GRB properties have been proposed in the literatures, and they can be classified into prompt correlations, afterglow correlations and prompt-afterglow correlations based on the episode in which the observables are measured \citep{2017NewAR..77...23D,2018PASP..130e1001D}. Among the prompt correlations, the so-called ``Amati relation'' \citep{2002AA...390...81A} has been widely discussed. It describes a relation between the isotropic-equivalent energy ($E_{\rm iso}$) and the rest-frame peak energy ($E_{\rm peak}$) of the $\gamma$-ray spectrum. Another commonly cited relation is the ``Yonetoku relation'' \citep{2003MNRAS.345..743W,2004ApJ...609..935Y}, which demonstrates the connection between $E_{\rm peak}$ and the isotropic–equivalent luminosity ($L_{\rm iso}$). The interpretation of these relations are still under debate \citep{2006ApJ...645L.113C,2006MNRAS.372.1699G}, and it is noteworthy that though the Amati relation and Yonetoku relation exist in both LGRBs and SGRBs, the best-fit parameters of the two relations are very likey to be different for the two GRB categories \citep{2012ApJ...750...88Z}. Since LGRBs and SGRBs are thought to have different origins but may share similar jet launching and radiation mechanisms, the correlations may help to discriminate among different models. 

The prompt correlations related to the intrinsic energy/luminosity can serve as distance indicators, hence can help to study the luminosity function and redshift distribution of GRBs \citep{2002AA...390...81A,2003A&A...407L...1A,2004ApJ...612L.101D,2004ApJ...609..935Y,2005ApJ...633..611L,2014ApJ...789...65Y,2018ApJ...852....1Z,2018AdAst2018E...1D,2019gbcc.book.....D}. However the large dispersion, which is about an order of magnitude for both Amati relation and  Yonetoku relation brings extra uncertainties on such studies, hence finding new correlations as redshift/distance indicators is more important and challenging.

Inspired by the $L_{\rm iso}-E_{\rm peak}-T_{0.45}$ relation for long GRBs found by \citet{2006MNRAS.370..185F}, here we manage to find a universal correlation for both LGRBs and SGRBs using the prompt emission properties $E_{\rm iso}$, $L_{\rm iso}$, $E_{\rm peak}$ and $T_{0.45}$. This paper is arranged as follows. In Section \ref{data}, we describe the selection of GRB samples; in Section \ref{correlations} we analyze the reliability of some potential correlations, and report the discovery of a universal correlation; in Section \ref{application}, the application of the universal correlation on the study of SGRB's luminosity function and formation rate is presented; finally, we give conclusions and discussion in Section \ref{SD}.  

\section{Sample Selections}\label{data}

In order to reduce the uncertainties in the correlation study, we require a careful selection on the GRBs samples. For the purpose of this work, we only include samples with the following information:
\begin{enumerate}
    \item the redshit z.
    \item the peak flux P and the peak fluence F.
    \item the peak energy $E_{\rm peak}^{\rm obs}$ in the observer's frame. And we take $E_{\rm peak} \equiv (1+z)E_{\rm peak}^{\rm obs}$ as the cosmological rest-frame $\nu f_\nu$ spectrum peak energy (in brief, the rest-frame peak energy).
    \item low-energy power-law index $\alpha$ and high-energy power-law index $\beta$ of Band fuction \citep{1993ApJ...413..281B}.
    \item $T_{0.45}^{\rm obs}$, and we take $T_{0.45} \equiv T_{0.45}^{\rm obs}/(1+z)$.
\end{enumerate}

One uncertainty for calculating a burst's intrinsic energy/luminosity comes from the assumed spectrum in the integration, so the 4th criterion above is to ensure we use a unified spectral form to evaluate $E_{\rm iso}$ and $L_{\rm iso}$, and the most energetic part of the emission is within the observed frequency band. As a consequence, all of our samples are reported to be best fitted with the Band function:
\begin{equation}
  \mathop N\nolimits_{{\rm{band}}}  = \left\{ {\begin{array}{*{20}{c}}
  {A{{\left( {E/100} \right)}^\alpha }\exp \left( { - E\left( {2 + \alpha } \right)/{E_{{\rm{peak}}}}} \right)}&{{\rm{if\ }}E < {E_{\rm{b}}}}\\
  {A{{\left\{ {\left( {\alpha  - \beta } \right){E_{{\rm{peak}}}}/\left[ {100\left( {2 + \alpha } \right)} \right]} \right\}}^{\left( {\alpha  - \beta } \right)}}\exp \left( {\beta  - \alpha } \right){{\left( {E/100} \right)}^\beta }}&{{\rm{if\ }}E \ge {E_{\rm{b}}}}
\end{array}} \right.
\end{equation}
where ${E_{\rm{b}}} = ({\alpha  - \beta }){E_{{\rm{peak}}}}/( {2 + \alpha })$. Finally, after applying all the selection criteria, our sample includes 49 LGRBs and 19 SGRBs. In Table \ref{tab:1} we list the information of the 49 LGRBs, and the information of the selected SGRBs are listed in Table \ref{tab:2}. In our following analysis we also include the 20 LGRB samples in \citet{2006MNRAS.370..185F}. 

In order to eliminate the influence of different observational energy bands on the calculating results, we make K-correction \citep{2001AJ....121.2879B} in the calculation of bursts' isotropic-equivalent luminosity/energy:
\begin{equation}\label{liso}
L_{\rm iso} = 4\pi D_{\rm L}^2(z)PK 
\end{equation}
\begin{equation}
d_{\rm L}(z)=\frac{c}{H_0}\int_{0}^{z}\frac{dz}{\sqrt{1-\Omega_{\rm m}+\Omega_{\rm m}(1+z)^3 }} 
\end{equation}
\begin{equation}
K=\frac{\int_{1KeV}^{10^4KeV}Ef(E)dE}{\int_{E_{\rm
 min}(1+z)}^{E_{\rm max}(1+z)}Ef(E)dE}
\end{equation}
where $P$ is the peak flux observed between a certain energy range $(E_{\rm min},E_{\rm max})$ in the unit of $\rm erg/cm^2/s$ (For $E_{\rm iso}$, $P$ is replaced by the Fluence $F$ in Eq.(\ref{liso})). We assume a flat $\Lambda$ cold dark matter universe with $\Omega_{\rm m}=0.27$ and $H_0=70\ \rm{km\ s^{-1}Mpc^{-1}}$ in the calculation. 

\section{Study on The Correlations}\label{correlations}
\subsection{The $E_{\rm iso}-E_{\rm peak}$ and $L_{\rm iso}-E_{\rm peak}$ Correlations}

In previous studies, a plenty of works about LGRB's prompt correlations were carried out. For SGRBs, however, such studies are much less due to the number of SGRBs with konwn redshifts are very small. Nevertheless, several studies have shown that for SGRBs they may follow a different $E_{\rm iso}-E_{\rm peak}$ relation with respect to LGRBs, while the $L_{\rm iso}-E_{\rm peak}$ relation may be similar to that of  LGRBs \citep{2012ApJ...755...55Z,2013MNRAS.430..163Q,2012ApJ...755...55Z,2014ApJ...789...65Y,2014PASJ...66...42T}. In the following we will test these correlations with our samples (49 LGRBs and 19 SGRBs together with 20 LGRBs in Firmani et al. (2006)).

\begin{figure}[ht!]
	\figurenum{1}\label{fig:fit1}
	\centering
	\includegraphics[angle=0,scale=0.4]{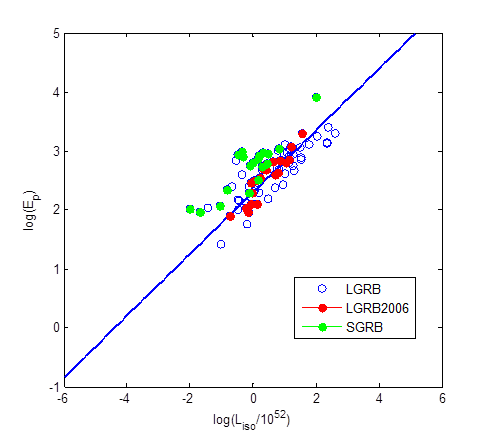}
	\includegraphics[angle=0,scale=0.4]{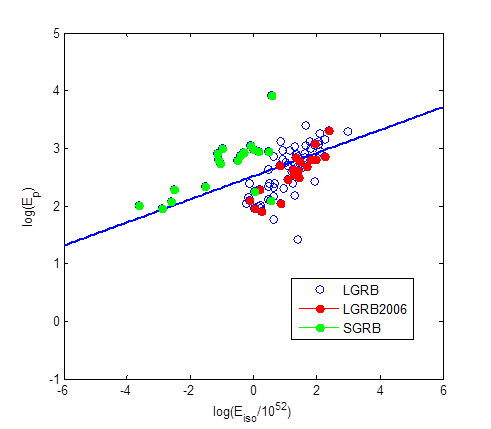}
	\caption{The $E_{\rm iso}-E_{\rm peak}$ and $L_{\rm iso}-E_{\rm peak}$ for LGRBs and SGRBs, the blue hollow points is the data for LGRBs in table 1, the red solid points is the data for LGRBs in \citet{2006MNRAS.370..185F} , the green solid points is the data for SGRBs in table 2. The solid line is the best-fit line for LGRBs and SGRBs. The Spearman’s rank correlation coefficients are P= 0.71 and P= 0.46 for left and right, respectively.}
	\hfill
\end{figure}

Here we constructed a likelihood function \citep{ 2006ApJ...653.1049W,2007ApJ...665.1489K} to fit the data. It can be defined as:
\begin{equation}\label{the_L}
\L(a_{j},b|x_{j},y)=\prod_{i}\frac{1}{\sqrt{2\pi(\sigma_{yi}^2+\sum_{j}(a_{j}\sigma_{xji})^2)}}exp({-\frac{1}{2}\frac{(y_{i}-b-\sum_{j}a_{i}x_{ji})^2}{\sigma_{yi}^2+\sum_{j}(a_{j}\sigma_{xji})^2}})
\end{equation}
where $y_i$ and $x_{ji}$ are the $E_{\rm peak}$ and $E_{\rm iso}$(or $L_{\rm iso}$) of GRB samples, $\sigma_{yi}$ and $\sigma_{xji}$ are their measurement errors. As shown by \citet{2002ApJ...574..740T} and \citet{2006ApJ...653.1049W}, the above likelihood function can be expressed as:
\begin{equation}\label{the_log(L)}
\log[L(a_{j},b|x_{j},y)]=- \sum_{i}\frac{(y_i-b-\sum_{j}a_{j}x_{ji})^2}{\sigma_{yi}^2+\sum_{j}(a_{j}\sigma_{xji})^2}+constant=-\chi^2+constant
\end{equation}
where $\chi^2$ is merit function \citep{1992nrca.book.....P,2003MNRAS.345.1057M,2006A&A...456..439K}. We determine the values of the parameters and their confidence intervals by performing Bayesian estimation based on the likelihood constructed above, we set uniform priors for all parameters. The results are presented in Fig.\ref{fig:fit1}. Obviously, the SGRB samples show systematically bias against the best-fit curve (see Section \ref{corrlation} for more details), thus it is inadequate to fit both LGRB and SGRB samples with a single relation on $E_{\rm iso}-E_{\rm peak}$ or $L_{\rm iso}-E_{\rm peak}$ plane.

\subsection{Discovery of a Universal Three-Parameter Correlation}\label{corrlation}

\citet{2006MNRAS.370..185F} found a correlation between $L_{\rm iso}$, $E_{\rm peak}$ and $T_{0.45}$ of LGRBs. Inspired by their work, and benefit from the accumulating number of SGRBs with know redshifts these years, we may be able to clarify whether both the SGRBs and LGRBs have the same three-parameter correlation. In the following we discuss both $L_{\rm iso}-E_{\rm peak}-T_{0.45}$ and $E_{\rm iso}-E_{\rm peak}-T_{0.45}$ relations.  

\begin{figure}[ht!]
	\figurenum{2}\label{fig:fit2}
	\centering
	\includegraphics[angle=0,scale=0.4]{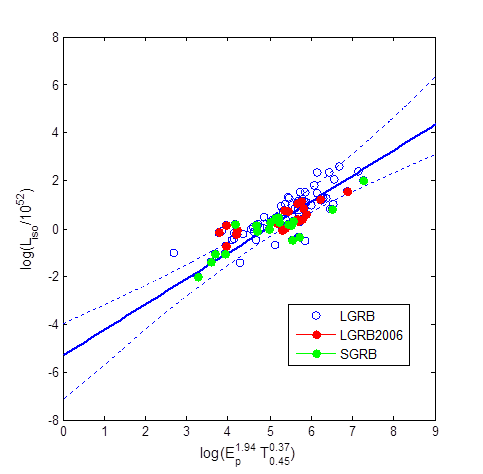}
	\includegraphics[angle=0,scale=0.4]{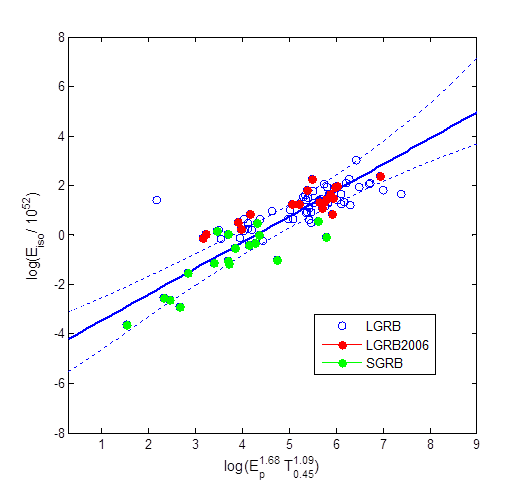}
	\caption{The three-parameter relations for LGRBs and SGRBs, the blue hollow points is the data for LGRBs in table 1, the red solid points is the data for LGRBs in \citet{2006MNRAS.370..185F}, the green solid points is the data for SGRBs in table 2. The solid line is the best fit line. The dotted lines represent  the 3$\sigma$ confidence bands. The Spearman’s rank correlation coefficients are P= 0.81 and P= 0.78 for left and right, respectively.}
	\hfill
\end{figure}

Considering the relations have the form of $L_{\rm iso} \propto E_{\rm peak}^{p1} T_{0.45}^{p2}$ ($E_{\rm iso} \propto E_{\rm peak}^{p1} T_{0.45}^{p2}$), and using the statistics defined by Eq.(\ref{the_log(L)}), we derive the best-fit correlation:
\begin{equation}\label{the_relation}
L_{\rm iso} \propto E_{\rm peak}^{1.94 \pm 0.06} T_{0.45}^{0.37 \pm 0.11}
\end{equation}
\begin{equation}
E_{\rm iso} \propto E_{\rm peak}^{1.68 \pm 0.09} T_{0.45}^{1.09 \pm 0.13}
\end{equation}
Where the errors are reported in 1 $\sigma$ confidence level. The results are shown in Fig.\ref{fig:fit2} together with their $3\sigma$ intervals. 

\begin{figure}[ht!]
	\figurenum{3}\label{fig:res}
	\centering
	\includegraphics[angle=0,scale=0.8]{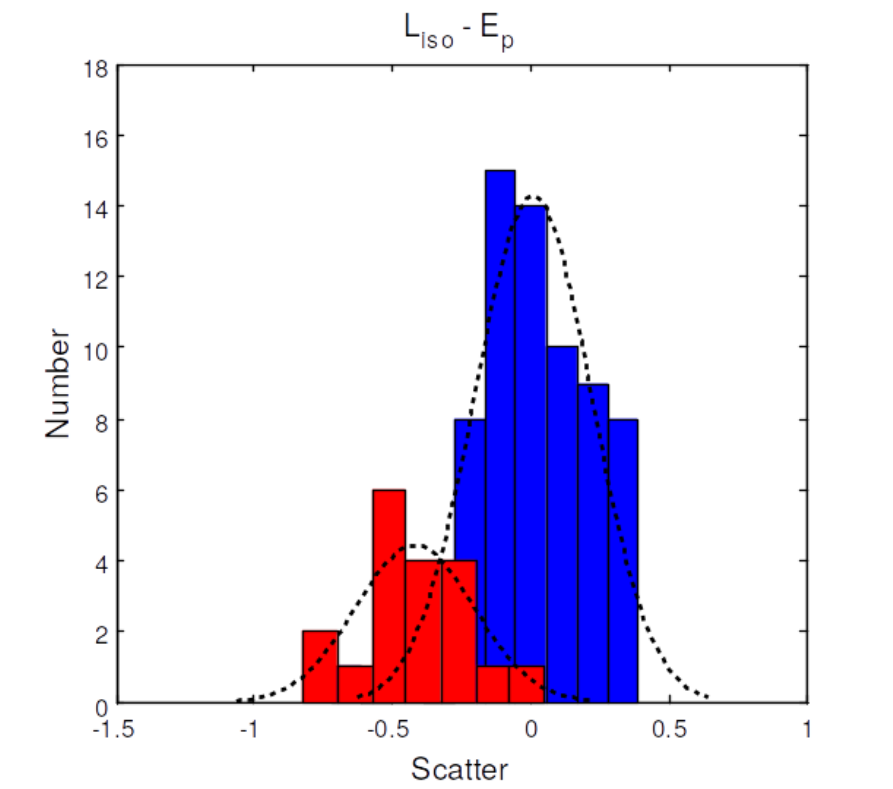}
	\includegraphics[angle=0,scale=0.8]{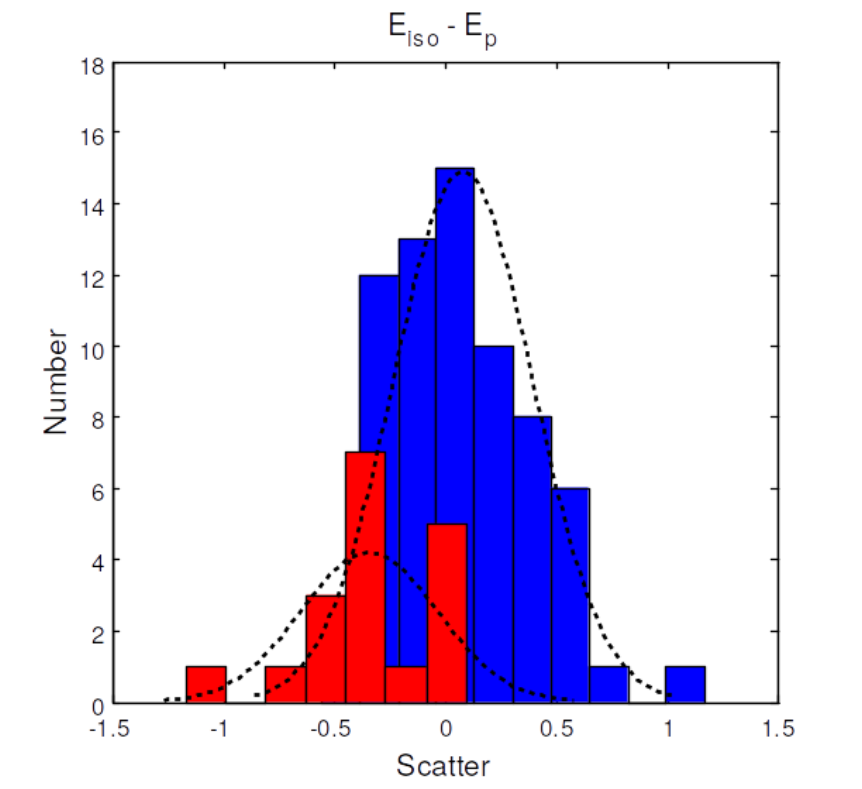}
	\includegraphics[angle=0,scale=0.8]{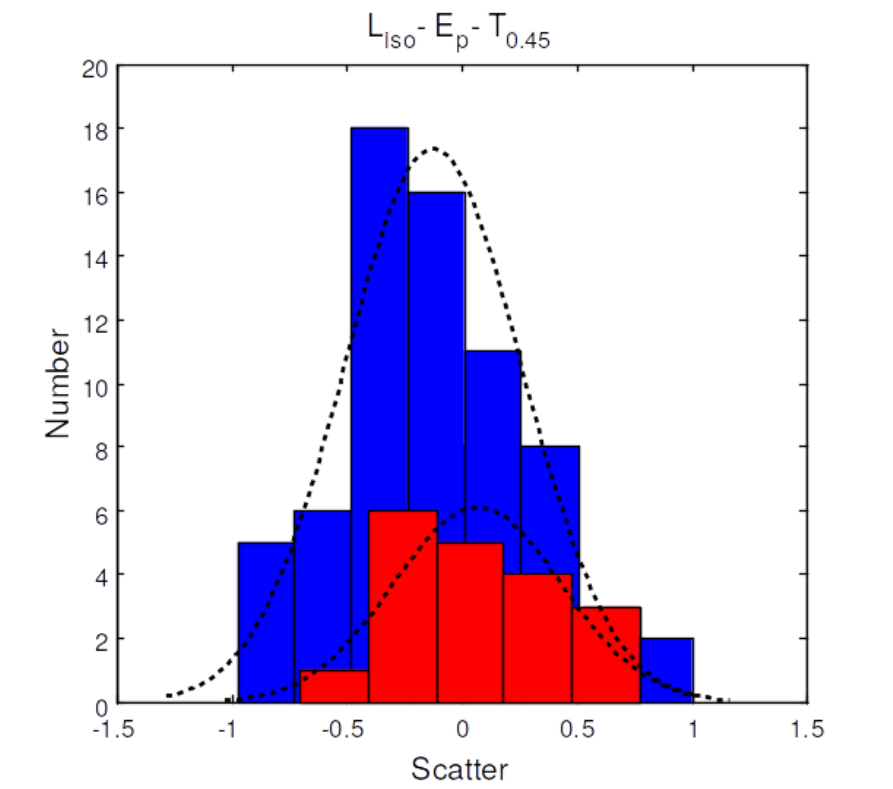}
	\includegraphics[angle=0,scale=0.8]{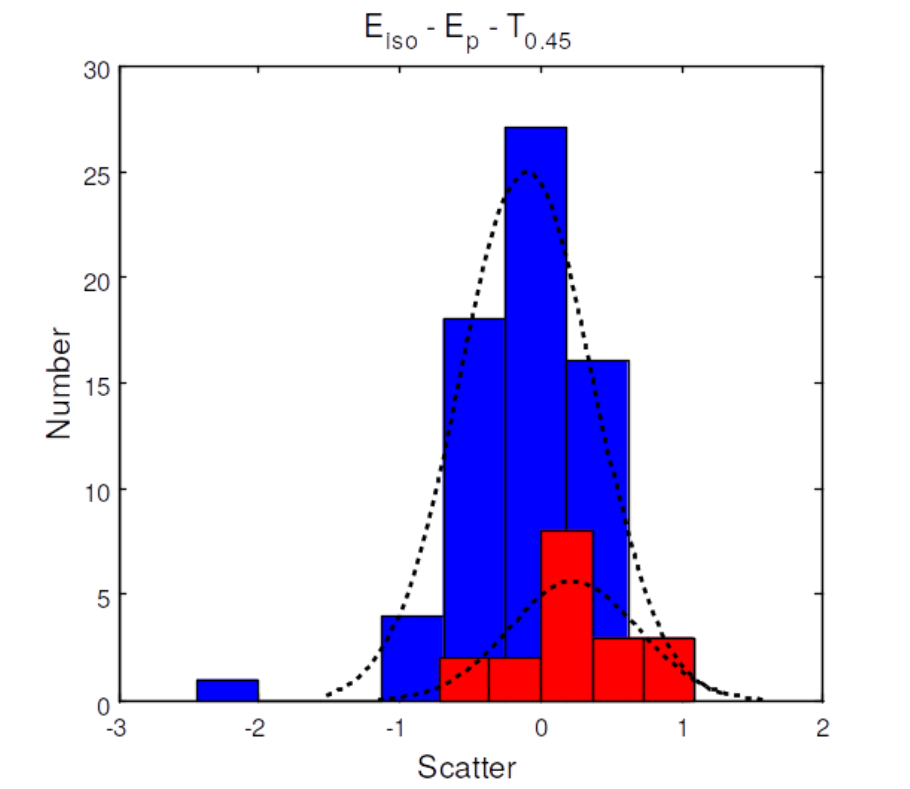}
	\caption{Left: The graph at the top shows that scatter with gaussian distribution of LGRB and SGRB for $L_{\rm iso}-E_{\rm peak}$ relationship. The red part is the SGRBs' scatter with $\mu=-0.42$, $\sigma=0.21$, the blue part is the LGRBs' scatter with $\mu=0.01$, $\sigma=0.22$. The graph below shows that scatter of LGRB and SGRB for $L_{\rm iso}-E_{\rm peak}-T_{0.45}$ relationship. The red part is the SGRBs' scatter with $\mu=0.06$, $\sigma=0.37$, the blue part is the LGRBs' scatter with $\mu=-0.12$, $\sigma=0.38$. Right: The graph at the top shows that scatter with gaussian distribution of LGRB and SGRB for $E_{\rm iso}-E_{\rm peak}$ relationship. The red part is the SGRBs' scatter with $\mu=-0.34$, $\sigma=0.30$, the blue part is the LGRBs' scatter with $\mu=0.08$, $\sigma=0.31$. The graph below shows that scatter of LGRB and SGRB for $E_{\rm iso}-E_{\rm peak}-T_{0.45}$ relationship. The red part is the SGRBs' scatter with $\mu=0.23$, $\sigma=0.46$, the blue part is the LGRBs' scatter with $\mu=-0.10$, $\sigma=0.47$.}
	\hfill
\end{figure}

We quantitatively discuss the reliability of the relations above (i.e., $E_{\rm iso}-E_{\rm peak}$, $L_{\rm iso}-E_{\rm peak}$, $E_{\rm iso}-E_{\rm peak}-T_{0.45}$ and $L_{\rm iso}-E_{\rm peak}-T_{0.45}$) in two aspects. First, by using the Anderson-Darling test, we find that the residuals of the data points with respect to the best-fit line can be fitted with a Gaussian function. The residuals of each fit are plotted in Fig.\ref{fig:res}, and we fit the LGRB samples (blue) and SGRB samples (red) separately with Gaussian function. We find that in the $L_{\rm iso}-E_{\rm peak}-T_{0.45}$ and $E_{\rm iso}-E_{\rm peak}-T_{0.45}$ fits, the mean values of the residuals of the two GRB categories are consistent with zero within the error range, and their standard deviation ($\sigma$) of LGRB and SGRB also consist with each other, indicating that in this two correlations the dispersions of LGRB and SGRB  are nearly the same. We further perform Student's t test on the residuals, the null hypothesis is that the mean values of the residuals of LGRBs and SGRBs are equal. We find that, given the significant level $a = 0.01$, only the residuals in the $L_{\rm iso}-E_{\rm peak}-T_{0.45}$ fit pass the test, with $P=0.07$, while in other correlations, $P=0.009, 2\times 10^{-6}$ and $1.5\times10^{-11}$ for  $E_{\rm iso}-E_{\rm peak}-T_{0.45}$, $E_{\rm iso}-E_{\rm peak}$ and $L_{\rm iso}-E_{\rm peak}$ respectively. Second, the robustness of a relation can be reflected by the correlation coefficient, here we use the Spearman’s rank correlation coefficient. For the correlations $L_{\rm iso}-E_{\rm peak}-T_{0.45}$, $E_{\rm iso}-E_{\rm peak}-T_{0.45}$, $L_{\rm iso}-E_{\rm peak}$ and $E_{\rm iso}-E_{\rm peak}$, the correlation coeffiicients are 0.81, 0.78, 0.71 and 0.47 respectively, which indicates that the $L_{\rm iso}-E_{\rm peak}-T_{0.45}$ relation is tighter than any of other relations.  

Based on the analysis above, we conclude that LGRBs and SGRBs could follow a universal $L_{\rm iso}-E_{\rm peak}-T_{0.45}$ correlation as shown in Eq.(\ref{the_relation}). We also compare this relation with the Yonetoku relation which using only SGRB samples and find that this relation is also slightly tighter than SGRB's Yonetoku relation, therefore the  $L_{\rm iso}-E_{\rm peak}-T_{0.45}$ correlation can serve as a potential redshift indicator for SGRBs.  

\section{Application on the Study of Luminosity Function and Formation Rate of SGRBs}\label{application}
\subsection{A New Redshift Indicator}

\begin{figure}[ht!]
	\figurenum{4}\label{fig:redshift}
	\centering
	\includegraphics[angle=0,scale=0.9]{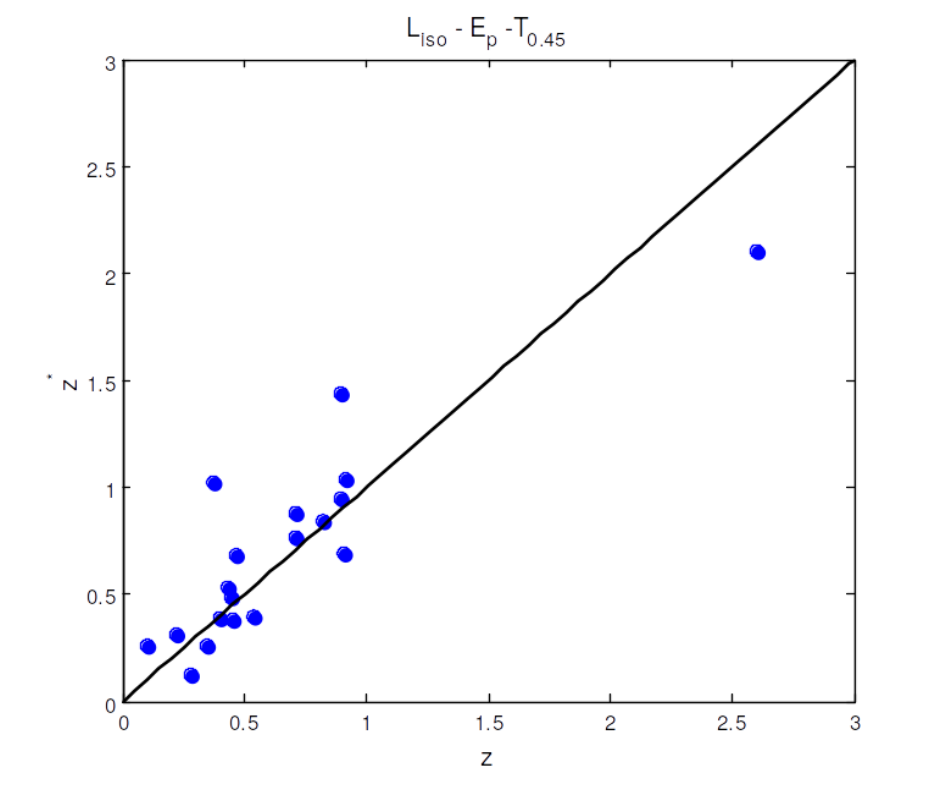}
	\caption{Comparison of real and pseudo-redshifts of 19 short GRBs. Y-axis is pseudo-redshifts, X-axis is real redshifts.}
	\hfill
\end{figure}

The $E_{\rm iso}-E_{\rm peak}$ and $L_{\rm iso}-E_{\rm peak}$ correlations were widely used to estimate the redshifts of GRBs \citep{2002AA...390...81A,2004ApJ...612L.101D,2004ApJ...609..935Y,2005ApJ...633..611L,2011MNRAS.418.2202D,2012int..workE.116A,2014ApJ...789...65Y,2018ApJ...852....1Z}. In order to apply the  $L_{\rm iso}-E_{\rm peak}-T_{0.45}$ correlation to the estimation of the redshift of GRBs, we first need to discuss the accuracy of the redshift calculated from this correlation. In Fig.\ref{fig:redshift}, we present the pseudo-redshifts $z^*$ calculated by the $L_{\rm iso}-E_{\rm peak}-T_{0.45}$ correlation versus the true observed redshift $z$ of our SGRB samples. By calculating the relative errors between pseudo-redshifts of SGRBs and observed redshifts, we find that the $68\%$ of our pseudo redshifts (i.e. 13 out of 19 sources) are within the $30\%$ of the real values. Then we perform the Kolmogorov–Smirnov test between them and find the chance probabilities of this is 0.74, which indicates that our estimated pseudo redshifts are reasonable. Then this encourages us to take a further step of using it in the study of Luminosity Function and Formation Rate of SGRBs.

\subsection{Constraining the Luminosity Function and Formation Rate of SGRBs}

We use the $L_{\rm iso}-E_{\rm peak}-T_{0.45}$ correlation to determine redshifts of SGRBs.  Our samples are collected from the \textit{Swift} GRB catalog with known $T_{0.45}$, peak flux and $E_{\rm peak}$ (or Bayes $E_{\rm peak}$ \footnote{\url{http://butler.lab.asu.edu/Swift/bat_spec_table.html}}) but lack of redshift detection. There are 75 bursts satisfy this criterion. Together with the 19 short burst samples with known redshifts, we use a total number of 94 bursts to study the luminosity function of SGRBs.

The observed GRB distributions suffer from selection effects, which are dominated by the Malmquist bias caused by the limited sensitivity of instruments. We use Lynden-Bell’s $c^-$ method, which has been widely used in the previous studies, to eliminate this effect \citep{1971MNRAS.155...95L,1992ApJ...399..345E,1993ApJ...402L..33P,1999ApJ...518...32M,2002ApJ...574..554L,2004ApJ...609..935Y,2014ApJ...789...65Y,2012MNRAS.423.2627W,2013ApJ...774..157D,2015ApJS..218...13Y,2015ApJ...800...31D,2016A&A...587A..40P, 2017ApJ...850..161T,2018ApJ...852....1Z}, The distribution of SGRB can be written as $\Psi(L,z)=\psi(L)\rho(z)$ \citep{1992ApJ...399..345E}, where $\psi(L)$ is the luminosity function and $\rho(z)$ is the formation rate of SGRBs. However in general the luminosity and redshift is not independent \citep{2002ApJ...574..554L}, the luminosity function $\psi(L)$ could still evolve with redshift z, and this degeneracy can be eliminated by adjusting the luminosity L with a factor $g(z)$, so that $\Psi(L,z)=\rho(z)\psi(L/g(z))/g(z)$, where $g(z)$ means the luminosity evolution, and the value $L_0=L/g(z)$ corresponds to the luminosity after removing the luminosity evolution effect. By making such substitution, the $\psi(L/g(z))$ is independent of redshift and represents the local luminosity distribution.

$g(z)$ is often taken as $g(z)=(1+z)^k$ in the literatures \citep{2002ApJ...574..554L,2014ApJ...789...65Y,2015ApJS..218...13Y}. Following \citet{1992ApJ...399..345E}, we use the non-parametric test method of $\tau$ statistical to derive the value of $k$. We get $k=4.78^{+0.17}_{-0.18}$ (the error is reported in 1$\sigma$ confidence level). For comparison, \citet{2014ApJ...789...65Y} gave $k=3.3^{+1.7}_{-3.7}$ using Swift samples, \citet{2018MNRAS.477.4275P} gave $k=4.269^{+0.134}_{-0.134}$ and \citet{2018ApJ...852....1Z} gave $k=4.47^{+0.47}_{-0.47}$ for Fermi samples.

\begin{figure}[ht!]
	\figurenum{5}\label{fig:distribution}
	\centering
	\includegraphics[angle=0,scale=0.26]{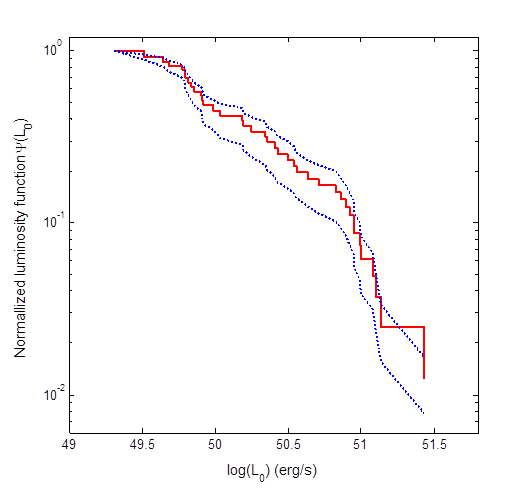}
	\includegraphics[angle=0,scale=0.26]{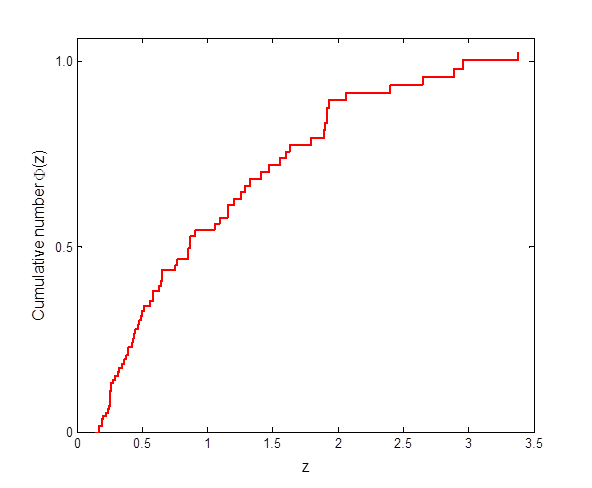}
	\includegraphics[angle=0,scale=0.26]{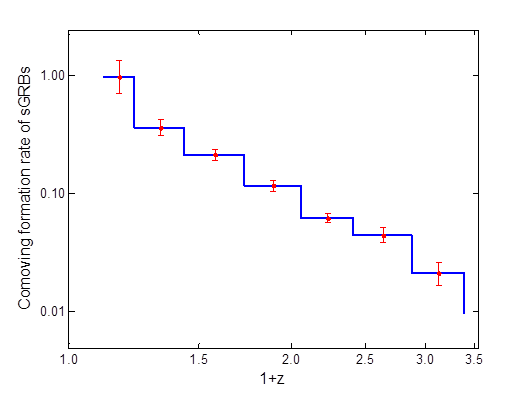}
	\caption{Left: cumulative luminosity function of SGRBs. The dotted lines are the $95\%$ confidence bands (Moreira et al. 2010); middle: cumulative redshift distribution of SGRBs; right: the PDF of the redshift distribution derived from the cumulative distribution, with its first bin normalized to unity.}
	\hfill
\end{figure}

After removing the effect of luminosity evolution through $L_0=L/(1+z)^{4.78}$, the cumulative luminosity function can be obtained by the following method \citep{1971MNRAS.155...95L,1992ApJ...399..345E}. For each point $(L_i, z_i)$, we define the set $J_i$ as
\begin{equation}
J_i=\lbrace{j}\mid{L_j}\geq {L_i},{z_j}\leq {z^{max}_{i}}\rbrace
\end{equation}
where $L_i$ is the luminosity of $i$th SGRB, the parameter $z^{max}_{i}$ is the maximum redshift at which the SGRBs with the luminosity $L_i$ can be detected. The number of SGRBs contained in this region is $n_i$. Then we use the following equation to calculate the cumulative luminosity function \citep{1971MNRAS.155...95L}
\begin{equation}
\psi(L_{0i})=\prod_{j<i}(1+\frac{1}{N_j})
\end{equation}
where $j < i$ means that the $j$th SGRB has a larger luminosity than $i$th sGRB. The results are shown in the left panel of Fig.\ref{fig:distribution}, which can be fitted with a broken power-law as
\begin{eqnarray}\label{lcmf}
\psi(L_0) \propto \left\{ 
\begin{array}{cc}
L_0^{-0.63 \pm 0.07} \qquad L_0 \le L_0^b \\
L_0^{-1.96 \pm 0.28} \qquad L_0 > L_0^b
\end{array} \right. 
\end{eqnarray}
where $L_0^b = 6.95_{-0.76}^{+0.84}\times10^{50} \rm{erg/s}$ is the break luminosity. This result is roughly in agreement with previous works \citep{2014ApJ...789...65Y, 2018ApJ...852....1Z}. For the formation rate of sGRBs, we define $J^{\prime}_i$  as
\begin{equation}
J_i^{\prime}=\lbrace{j}\mid{L_j}> {L_i^{lim}},{z_j}< {z_{i}}\rbrace
\end{equation}
Where $z_i$ is the redshift of $i$th SGRB, the parameter $L^{min}_i$ is the minimum luminosity which can be detected at redshift $z_i$. The number of SGRBs contained in this region is $m_i$. Then we can obtain the cumulative redshift distribution as \citep{1971MNRAS.155...95L}:
\begin{equation}
\phi(z_i)=\prod_{j<i}(1+\frac{1}{M_j})
\end{equation}
where $j<i$ means that the $j$th SGRB has a less redshift than $i$th SGRB. The results are shown in the middle panel of Fig.\ref{fig:distribution}. Then the probability density function (PDF) of redshift distribution can be calculated by:
\begin{equation}
\rho(z)=\frac{d\phi(z)}{dz}(1+z)(\frac{dV(z)}{dz})^{-1}
\end{equation}
\begin{equation}
\frac{dV(z)}{dz}=4\pi \left ( \frac{c}{H_0} \right )^3 \left [ \int_{0}^{z}\frac{dz}{\sqrt{1-\Omega_{\rm m}+\Omega_{\rm m}(1+z)^3}}\right ]^2 \times \frac{1}{\sqrt{1-\Omega_{\rm m}+\Omega_{\rm m}(1+z)^3}}
\end{equation}

The results are shown in the right panel of Fig.\ref{fig:distribution}. Again, the formation rate of SGRBs can be fitted by:
\begin{eqnarray}\label{zpdf}
\rho(z) \propto \left\{ 
\begin{array}{cc}
(1+z)^{-4.39 \pm 0.55} \qquad z \le 1.5 \\
(1+z)^{-5.51 \pm 0.32} \qquad z > 1.5
\end{array} \right. 
\end{eqnarray}

Meanwhile, we can derive the formation rate of sGRBs in the local universe, $\rho(0)=15.5 \pm 5.8\ \rm{Gpc^{-3}yr^{-1}}$, which is roughly consistent with the results of \citet{2015ApJ...815..102F} that the local event rate is $10 \rm{ Gpc^{-3}yr^{-1}}$ and $\rho(0)=7.53 \rm{Gpc^{-3}yr^{-1}}$  in \citet{2018ApJ...852....1Z}.

\begin{figure}[ht!]
	\figurenum{6}\label{fig:mctest}
	\centering
	\includegraphics[angle=0,scale=0.4]{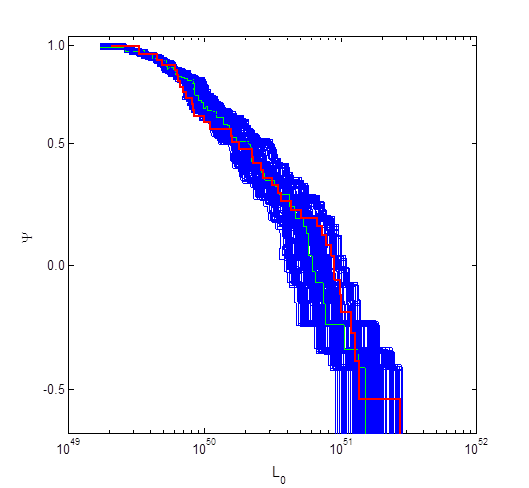}
	\includegraphics[angle=0,scale=0.4]{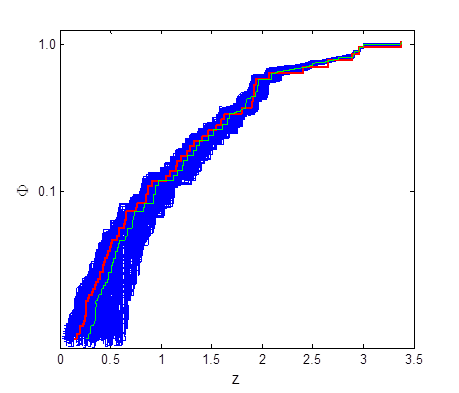}
	\caption{Comparison of simulated data (blue) and observed data (red). Left is the cumulative luminosity function. Right is the cumulative redshift distribution. The green lines are the mean of simulated data.}
	\hfill
\end{figure}

At last, we perform the Monte Carlo simulation to test whether our results can recover the sGRB distributions. We simulate a set of points $(L_0, z)$ which follow the equations Eq.(\ref{lcmf}) and Eq.(\ref{zpdf}), and then calculate the luminosity $L$ using the relation $L=L_0(1+z)^{4.78}$, thus we obtain a set of points $(L, z)$. We simulate $10^{5}$ points and divide them into 100 groups. For each group, we select one sGRBs from them and get 100 pseudo sGRBs to compare with the observed data. Finally, we perform the same analysis as above to obtain the luminosity function and formation rate of sGRBs. In Fig.\ref{fig:mctest} we present our results, where the blue lines are the simulated data for the cumulative luminosity function and cumulative redshift distribution, the red lines and green lines are the observed data and the mean of the simulated data, respectively. We also perform the Kolmogorov–Smirnov test between observed data and the mean distribution of simulated data, the chance probabilities are 0.62 and 0.83 respectively, which indicates that the cumulative luminosity function and formation rate are reliable.

\section{Summary and Discussion}\label{SD}

Gamma-ray bursts are the most powerful explosions in the universe, although great progress have been achieved in recent years, their nature is still unclear. Meanwhile since their huge luminosity/energy, GRBs can be detected to large redshift and thus can serve as a valuable cosmological tool.

A variety of correlations among intrinsic properties have been proposed in these years, on one hand these correlations can help to reveal the nature of GRBs, and on the other hand some correlations may be used as a redshift indicator, such as the so-called “Amati relation” \citep{2002AA...390...81A} and “Yonetoku relation” \citep{2014ApJ...789...65Y}. However none of these correlations is valid for both long GRBs and short GRBs. In this paper, we study the correlations among gamma-ray burst prompt emission properties and find a universal correlation which is suitable for both long and short GRBs, i.e. $L_{iso}\propto {E^{1.94}_{peak}}T^{0.37}_{0.45}$ This universal correlation may have important implications for GRB physics, implying that the long and short GRBs may share similar radiation processes. The parameters of the relation obtained in our study is different from that in \citet{2006MNRAS.370..185F}. The reason may not only be the different sample (more long GRBs and the inclusion of short GRBs) we used in our study, but could also be the degeneracy between parameters in the three-parameter fitting. The degeneracy could be studied by examine the joint posterior distributions obtained via MCMC/nested sampling approaches, and we leave it for future study.

Some other three-parameter correlations have also been found for GRBs, for example, \citet{2005ApJ...633..611L} found that there is a tight correlation between the isotropic gamma-ray energy, the peak energy of the gamma-ray spectra, and the break time of the optical afterglow light curves. For GRBs with X-ray plateau phase, the three-parameter correlations also have been reported to exist between the end time of the plateau phase, the corresponding X-ray luminosity and the peak luminosity or isotropic energy in the prompt emission phase \citep{2014PASJ...66...42T,2016ApJ...825L..20D,2017ApJ...848...88D, 2019ApJS..245....1T}. In these relations we need to know the information of the afterglow emission, such as the break time of the afterglow light curves, the duration and  luminosity of the X-ray plateau phase. Nonetheless, the universal correlation found in this paper only involves properties of GRB prompt emission and does not require any information of afterglow phase.

This universal correlation can be used as a redshift indicator. Here we use this relation to calculate the pseudo redshifts of short GRBs, and then use the Lynden-Bell’s $c^-$ method to study the luminosity function and formation rate of SGRBs. We find that the luminosity function can be expressed as $\psi(L_0)\propto{L_0^{-0.63\pm{0.07}}}$ for dim SGRBs and $\psi(L_0)\propto{L_0^{-1.96\pm{0.28}}}$ for bright SGRBs, with the break luminosity $L^b_0=6.95_{-0.76}^{+0.84}\times{10^{50}} \rm{erg/s}$. For the formation rate of sGRBs, we give the result that $\rho(z)\propto(1+z)^{-4.39\pm{0.55}}$ for $z<1.5$ and $\rho(z)\propto(1+z)^{-5.51\pm{0.32}}$ for $z>1.5$, and also we have obtained the local SGRB rate as $\rho(0)=15.5 \pm 5.8\ \rm{Gpc^{-3}yr^{-1}}$, these results are roughly consistent with \citet{2018ApJ...852....1Z}. If we take the typical beaming factor as $f_b^{-1}=30$ \citep{2015ApJ...815..102F}, then the actual total local event rate of SGRBs is about $450 \rm{Gpc^{-3}yr^{-1}}$, which is also consistent with the results inferred from the gravitational wave detections \citep{2019PhRvX...9c1040A}.

\acknowledgments

We thank the anonymous referee for valuable comments. We thank Jinyi Shangguan for helpful discussions. This work was supported by NSFC (No. 11933010), by the Chinese Academy of Sciences via the Strategic Priority Research Program (No. XDB23040000), Key Research Program of Frontier Sciences (No. QYZDJ-SSW-SYS024).

\clearpage

\begin{deluxetable*}{lrrrrrrrrl}
\tablecaption{The 49 Long GRB samples\label{tab:1}}
\tablewidth{700pt}
\tabletypesize{\scriptsize}
\tablehead{
\colhead{Name} & \colhead{$z$} & \colhead{$E_{\rm peak}$} &
\colhead{$T_{\rm 0.45}$} & \colhead{$\alpha$} & \colhead{$\beta$} &
\colhead{$L_{\rm iso}$} & \colhead{$E_{\rm iso}$}  & \colhead{Detection Band} & \colhead{Ref.} \\ 
\colhead{} & \colhead{} & \colhead{(KeV)} & 
\colhead{(s)} & \colhead{} & \colhead{} &
\colhead{($10^{52}$erg/s)} & \colhead{($10^{52}$erg/s)} & \colhead{(KeV)} & \colhead{} 
} 
\startdata
50401	&	2.9	&	$	464.10	\pm	26	$	&	$	5.18	\pm	0.212	$	&	-0.83	&	-2.37	&	$	20.90	\pm	0.1	$	&	$	35.00	\pm	7	$	&	$	20-2000	$	&	2	\\
050416A	&	0.6535	&	$	26.01	\pm	2	$	&	$	0.63	\pm	0.043	$	&	-1	&	-3.4	&	$	0.10	\pm	0.99	$	&	$	25.10	\pm	0.01	$	&	$	15-150	$	&	1	\\
50603	&	2.821	&	$	1333.53	\pm	28	$	&	$	1.6	\pm	0.08	$	&	-0.79	&	-2.15	&	$	225.00	\pm	0.14	$	&	$	60.00	\pm	4	$	&	$	20-3000	$	&	3	\\
60124	&	2.296	&	$	816.62	\pm	88	$	&	$	2.38	\pm	0.129	$	&	-1.29	&	-2.25	&	$	13.70	\pm	0.74	$	&	$	41.00	\pm	6	$	&	$	20-2000	$	&	4	\\
61007	&	1.261	&	$	1125.98	\pm	48	$	&	$	16.8	\pm	0.122	$	&	-0.53	&	-2.61	&	$	17.80	\pm	0.31	$	&	$	86.00	\pm	9	$	&	$	20-10000	$	&	5	\\
061222A	&	2.088	&	$	1090.06	\pm	54	$	&	$	8.97	\pm	0.159	$	&	-1	&	-2.32	&	$	14.80	\pm	0.4	$	&	$	20.80	\pm	0.602	$	&	$	20-10000	$	&	1	\\
071010B	&	0.947	&	$	101.24	\pm	10	$	&	$	4.68	\pm	0.075	$	&	-1.25	&	-2.65	&	$	0.65	\pm	2.17	$	&	$	1.70	\pm	0.9	$	&	$	20-1000	$	&	6	\\
080319B	&	0.937	&	$	1307.48	\pm	22	$	&	$	20.3	\pm	0.058	$	&	-0.86	&	-3.59	&	$	10.50	\pm	0.1	$	&	$	114.00	\pm	9	$	&	$	20-7000	$	&	1	\\
080319C	&	1.95	&	$	905.65	\pm	0.06	$	&	$	5.04	\pm	0.152	$	&	-1.01	&	-1.87	&	$	9.46	\pm	2.28	$	&	$	14.10	\pm	2.8	$	&	$	20-4000	$	&	7	\\
80605	&	1.6398	&	$	784.02	\pm	40	$	&	$	4.75	\pm	0.079	$	&	-0.87	&	-2.58	&	$	33.30	\pm	0.69	$	&	$	24.00	\pm	2	$	&	$	20-2000	$	&	8	\\
80607	&	3.036	&	$	1404.53	\pm	27	$	&	$	7.52	\pm	0.094	$	&	-0.76	&	-2.57	&	$	221.00	\pm	0.44	$	&	$	188.00	\pm	10	$	&	$	20-4000	$	&	9	\\
80721	&	2.602	&	$	1790.19	\pm	62	$	&	$	4.51	\pm	0.195	$	&	-0.96	&	-2.42	&	$	111.00	\pm	0.18	$	&	$	126.00	\pm	22	$	&	$	20-7000	$	&	1	\\
80810	&	3.35	&	$	2523.00	\pm	263	$	&	$	31.35	\pm	2.914	$	&	-1.2	&	-2.5	&	$	239.00	\pm	0.14	$	&	$	45.00	\pm	5	$	&	$	15-1000	$	&	11	\\
80913	&	6.7	&	$	931.70	\pm	39	$	&	$	2.16	\pm	0.139	$	&	-0.82	&	-2.5	&	$	12.40	\pm	0.18	$	&	$	8.60	\pm	2.5	$	&	$	15-150	$	&	12,13	\\
81028	&	3.038	&	$	240.91	\pm	6	$	&	$	78.12	\pm	2.09	$	&	0.36	&	-2.25	&	$	4.91	\pm	0.45	$	&	$	17.00	\pm	2	$	&	$	8-35000	$	&	10	\\
81121	&	2.512	&	$	726.63	\pm	43.8	$	&	$	5.51	\pm	0.26	$	&	-0.21	&	-1.86	&	$	13.80	\pm	0.22	$	&	$	26.00	\pm	5	$	&	$	8-35000	$	&	10	\\
81222	&	2.77	&	$	629.59	\pm	8	$	&	$	4.2	\pm	0.053	$	&	-0.9	&	-2.33	&	$	10.10	\pm	0.03	$	&	$	30.00	\pm	3	$	&	$	8-35000	$	&	1	\\
90424	&	0.544	&	$	250.13	\pm	2	$	&	$	1.98	\pm	0.03	$	&	-1.02	&	-3.26	&	$	1.14	\pm	0.02	$	&	$	4.60	\pm	0.9	$	&	$	8-35000	$	&	1	\\
90516	&	4.109	&	$	262.60	\pm	11.4	$	&	$	36.96	\pm	2.259	$	&	-1.03	&	-2.1	&	$	8.70	\pm	0.33	$	&	$	88.50	\pm	19.2	$	&	$	8-1000	$	&	14	\\
90618	&	0.54	&	$	482.33	\pm	14	$	&	$	23.04	\pm	0.706	$	&	-0.91	&	-2.42	&	$	1.87	\pm	0.08	$	&	$	25.40	\pm	0.6	$	&	$	8-35000	$	&	10	\\
90809	&	2.737	&	$	722.74	\pm	11	$	&	$	3.43	\pm	0.456	$	&	-0.47	&	-2.16	&	$	34.00	\pm	0.28	$	&	$	4.20	\pm	1.2	$	&	$	8-35000	$	&	10	\\
90927	&	1.37	&	$	141.42	\pm	1.81	$	&	$	0.84	\pm	0.09	$	&	-0.68	&	-2.12	&	$	0.37	\pm	0.23	$	&	$	0.70	\pm	0.312	$	&	$	8-35000	$	&	10	\\
91020	&	1.71	&	$	506.77	\pm	25	$	&	$	6	\pm	0.166	$	&	-1.2	&	-2.29	&	$	3.44	\pm	0.04	$	&	$	12.20	\pm	2.4	$	&	$	8-35000	$	&	1	\\
91127	&	0.49	&	$	88.91	\pm	1.81	$	&	$	1.54	\pm	0.063	$	&	-0.68	&	-2.12	&	$	0.77	\pm	0.19	$	&	$	1.63	\pm	0.02	$	&	$	8-35000	$	&	10	\\
100615A	&	1.398	&	$	205.58	\pm	8	$	&	$	9.43	\pm	0.222	$	&	-1.24	&	-2.27	&	$	1.06	\pm	0.03	$	&	$	4.22	\pm	1.21	$	&	$	8-1000	$	&	15	\\
100621A	&	0.542	&	$	146.49	\pm	15	$	&	$	19.43	\pm	0.327	$	&	-1.7	&	-2.45	&	$	0.32	\pm	0.25	$	&	$	4.37	\pm	0.5	$	&	$	20-2000	$	&	1	\\
100728A	&	1.567	&	$	1001.13	\pm	25	$	&	$	57.35	\pm	0.185	$	&	-0.47	&	-2.5	&	$	6.45	\pm	1.08	$	&	$	63.74	\pm	12.2	$	&	$	20-10000	$	&	16	\\
100816A	&	0.8049	&	$	246.73	\pm	4.7	$	&	$	0.84	\pm	0.016	$	&	-0.31	&	-2.77	&	$	0.74	\pm	0.12	$	&	$	0.73	\pm	0.02	$	&	$	10-1000	$	&	17,18	\\
100906A	&	1.727	&	$	490.86	\pm	40	$	&	$	11.07	\pm	0.123	$	&	-1.1	&	-2.2	&	$	6.90	\pm	0.77	$	&	$	28.90	\pm	0.3	$	&	$	20-2000	$	&	19	\\
101213A	&	0.414	&	$	437.91	\pm	40	$	&	$	20.16	\pm	0.69	$	&	-1.1	&	-2.35	&	$	0.06	\pm	0.43	$	&	$	3.01	\pm	0.64	$	&	$	10-1000	$	&	20	\\
101219B	&	0.55	&	$	108.50	\pm	8	$	&	$	8.36	\pm	0.736	$	&	-0.33	&	-2.12	&	$	0.04	\pm	0.38	$	&	$	0.59	\pm	0.04	$	&	$	10-1000	$	&	21	\\
110422A	&	1.77	&	$	681.42	\pm	34	$	&	$	9.24	\pm	0.09	$	&	-0.53	&	-2.65	&	$	0.29	\pm	0.36	$	&	$	43.10	\pm	0.13	$	&	$	20-2000	$	&	22	\\
110503A	&	1.613	&	$	572.25	\pm	19	$	&	$	2.1	\pm	0.056	$	&	-0.98	&	-2.7	&	$	18.90	\pm	0.19	$	&	$	10.49	\pm	11.4	$	&	$	20-5000	$	&	1	\\
110715A	&	0.82	&	$	218.40	\pm	11	$	&	$	1.45	\pm	0.025	$	&	-1.23	&	-2.7	&	$	1.19	\pm	0.39	$	&	$	2.93	\pm	0.12	$	&	$	20-10000	$	&	23	\\
110731A	&	2.83	&	$	1164.32	\pm	13	$	&	$	3.36	\pm	0.071	$	&	-0.8	&	-2.98	&	$	30.60	\pm	0.07	$	&	$	118.05	\pm	9.12	$	&	$	10-1000	$	&	24	\\
110801A	&	1.858	&	$	400.12	\pm	50	$	&	$	54.56	\pm	1.731	$	&	-1.7	&	-2.5	&	$	0.44	\pm	0.84	$	&	$	15.84	\pm	1.32	$	&	$	15-350	$	&	25	\\
111228A	&	0.714	&	$	58.28	\pm	3	$	&	$	9	\pm	0.307	$	&	-1.9	&	-2.7	&	$	0.67	\pm	0.25	$	&	$	4.41	\pm	0.202	$	&	$	10-1000	$	&	26	\\
120119A	&	1.728	&	$	516.14	\pm	8	$	&	$	13.6	\pm	0.198	$	&	-0.98	&	-2.36	&	$	5.98	\pm	0.14	$	&	$	20.79	\pm	1.98	$	&	$	10-1000	$	&	27	\\
120326A	&	1.798	&	$	129.97	\pm	3.67	$	&	$	4.68	\pm	0.114	$	&	-0.98	&	-2.53	&	$	0.59	\pm	0.1	$	&	$	3.11	\pm	0.617	$	&	$	10-1000 	$	&	28	\\
120712A	&	4.1745	&	$	641.64	\pm	26	$	&	$	4.68	\pm	0.146	$	&	-0.6	&	-1.8	&	$	13.50	\pm	0.08	$	&	$	8.35	\pm	1.49	$	&	$	10-1000	$	&	29	\\
120922A	&	3.1	&	$	154.57	\pm	3.5	$	&	$	43.93	\pm	1.488	$	&	-1.6	&	-2.3	&	$	3.02	\pm	0.27	$	&	$	19.79	\pm	5.89	$	&	$	10-1000	$	&	30	\\
121128A	&	2.2	&	$	199.04	\pm	4.6	$	&	$	4.75	\pm	0.198	$	&	-0.8	&	-2.41	&	$	1.53	\pm	0.19	$	&	$	9.24	\pm	1.11	$	&	$	10-1000	$	&	31	\\
130215A	&	0.597	&	$	247.54	\pm	63	$	&	$	19.5	\pm	1.148	$	&	-1	&	-1.6	&	$	0.22	\pm	0.18	$	&	$	3.14	\pm	0.88	$	&	$	10-1000	$	&	32	\\
130408A	&	3.758	&	$	1294.18	\pm	40	$	&	$	1.26	\pm	0.092	$	&	-0.7	&	-2.3	&	$	61.20	\pm	0.59	$	&	$	7.41	\pm	1.41	$	&	$	20-10000	$	&	33	\\
130427A	&	0.3399	&	$	1112.12	\pm	5	$	&	$	7.64	\pm	0.382	$	&	-0.789	&	-3.06	&	$	19.00	\pm	0.001	$	&	$	46.18	\pm	8.26	$	&	$	8-1000	$	&	34	\\
130505A	&	2.27	&	$	1975.08	\pm	49	$	&	$	6.23	\pm	0.623	$	&	-0.31	&	-2.26	&	$	398.00	\pm	0.17	$	&	$	1012.60	\pm	253.9	$	&	$	20-1200	$	&	35	\\
130831A	&	0.4791	&	$	99.10	\pm	4	$	&	$	4.07	\pm	0.094	$	&	-1.51	&	-2.8	&	$	0.34	\pm	0.39	$	&	$	1.49	\pm	0.0578	$	&	$	20-10000	$	&	36	\\
130907A	&	1.238	&	$	872.82	\pm	16	$	&	$	40.31	\pm	0.139	$	&	-0.65	&	-2.22	&	$	18.20	\pm	0.08	$	&	$	122.31	\pm	10.92	$	&	$	20-10000	$	&	37	\\
131030A	&	1.295	&	$	406.22	\pm	10	$	&	$	7.35	\pm	0.118	$	&	-0.71	&	-2.95	&	$	10.80	\pm	0.11	$	&	$	24.72	\pm	3.19	$	&	$	20-10000	$	&	38	\\
\enddata
\tablecomments{
References. (1) \citet{2012MNRAS.421.1256N}; (2) \citet{2005GCN..3179....1G}; (3) \citet{2005GCN..3518....1G}; (4) \citet{2006GCN..4599....1G}; (5) \citet{2006GCN..5722....1G}; (6) \citet{2007GCN..6879....1G}; (7) \citet{2008GCN..7487....1G}; (8) \citet{2008GCN..7854....1G}; (9) \citet{2008GCN..7862....1G}; (10) Nava et al. (2011); (11) \citet{2008GCN..8101....1S}; (12) \citet{2008GCN..8256....1P}; (13) \citet{2009ApJ...693.1610G}; (14) \citet{2009GCN..9415....1M}; (15) \citet{2010GCN.10851....1F}; (16) \citet{2010GCN.11021....1G}; (17) \citet{2010GCN.11124....1F}; (18) \citet{2010GCN.11124....1F}; (19) \citet{2010GCN.11251....1G}; (20) \citet{2010GCN.11454....1G}; (21) \citet{2010GCN.11477....1V}; (22) \citet{2011GCN.11971....1G}; (23) \citet{2011GCN.12166....1G}; (24) \citet{2011GCN.12223....1G}; (25) \citet{2011GCN.12276....1S}; (26) \citet{2011GCN.12744....1B}; (27) \citet{2012GCN.12874....1G}; (28) \citet{2012GCN.13145....1C}; (29) \citet{2012GCN.13469....1G}; (30) \citet{2012GCN.13809....1Y}; (31) \citet{2012GCN.14012....1M}; (32) \citet{2013GCN.14219....1Y}; (33) \citet{2013GCN.14368....1G} ; (34) \citet{2013GCN.14473....1V}; (35) \citet{2013GCN.14575....1G}; (36) \citet{2013GCN.15145....1G}; (37) \citet{2013GCN.15203....1G};  (38)  \citet{2013GCN.15413....1G}.
}
\end{deluxetable*}

\begin{deluxetable*}{lrrrrrl}
\tablecaption{The 19 Short GRB samples\label{tab:2}}
\tablewidth{700pt}
\tabletypesize{\scriptsize}
\tablehead{
\colhead{Name} & \colhead{$z$} & 
\colhead{$E_{\rm peak}$} & \colhead{$T_{\rm 0.45}$} & 
\colhead{$L_{\rm iso}$} & \colhead{$E_{\rm iso}$} & \colhead{Ref.} \\ 
\colhead{} & \colhead{} & \colhead{(KeV)} & \colhead{(s)} & 
\colhead{($10^{52}$erg/s)} & \colhead{($10^{52}$erg/s)} & \colhead{} 
} 
\startdata
050509B	&	0.225	&	$	102	\pm	10	$	&	$	0.02	\pm	0.04	$	&	$	0.01	\pm	0.09	$	&	$	0.00024	\pm	0.00044	$	&	14	\\
051221A	&	0.546	&	$	621	\pm	114	$	&	$	0.16	\pm	0.008	$	&	$	2.77	\pm	0.29	$	&	$	0.3	\pm	0.04	$	&	1,2	\\
060502B	&	0.287	&	$	193	\pm	19	$	&	$	0.04	\pm	0.007	$	&	$	0.089	\pm	0.05	$	&	$	0.003	\pm	0.005	$	&	1	\\
61201	&	0.111	&	$	969	\pm	508	$	&	$	0.22	\pm	0.014	$	&	$	0.3445	\pm	0.4	$	&	$	3	\pm	4	$	&	1	\\
61217	&	0.827	&	$	216	\pm	22	$	&	$	0.1	\pm	0.008	$	&	$	1.498	\pm	2.2	$	&	$	0.03	\pm	0.04	$	&	1	\\
070429B	&	0.902	&	$	813	\pm	81	$	&	$	0.08	\pm	0.011	$	&	$	1.873	\pm	1.6	$	&	$	0.07	\pm	0.11	$	&	1	\\
070714B	&	0.92	&	$	2150	\pm	1113	$	&	$	1.4	\pm	0.132	$	&	$	6.56	\pm	1.36	$	&	$	0.83	\pm	0.1	$	&	4	\\
070724A	&	0.457	&	$	119	\pm	12	$	&	$	0.11	\pm	0.011	$	&	$	0.087	\pm	0.005	$	&	$	0.00245	\pm	0.00175	$	&	13	\\
70809	&	0.473	&	$	91	\pm	9	$	&	$	0.26	\pm	0.018	$	&	$	0.042	\pm	0.001	$	&	$	0.00131	\pm	0.00103	$	&	1	\\
90510	&	0.903	&	$	8370	\pm	760	$	&	$	0.12	\pm	0.013	$	&	$	104	\pm	24	$	&	$	3.75	\pm	0.25	$	&	5	\\
090426A	&	2.609	&	$	320	\pm	54	$	&	$	0.32	\pm	0.025	$	&	$	1.46	\pm	0.38	$	&	$	1.1	\pm	0.38	$	&	1	\\
100117A	&	0.915	&	$	551	\pm	135	$	&	$	0.14	\pm	1.693	$	&	$	1.89	\pm	0.21	$	&	$	0.09	\pm	0.01	$	&	5	\\
100206A	&	0.407	&	$	638.98	\pm	131.21	$	&	$	0.06	\pm	0.009	$	&	$	1	\pm	1.15	$	&	$	0.0763	\pm	0.03	$	&	5	\\
100625A	&	0.452	&	$	1018	\pm	166	$	&	$	0.14	\pm	0.007	$	&	$	1.4	\pm	0.06	$	&	$	0.399	\pm	0.06	$	&	11	\\
101219A	&	0.718	&	$	842	\pm	170	$	&	$	0.24	\pm	0.011	$	&	$	1.56	\pm	0.24	$	&	$	0.49	\pm	0.23	$	&	6	\\
130603B	&	0.356	&	$	891.66	\pm	135.6	$	&	$	0.04	\pm	0.007	$	&	$	3.04	\pm	0.44	$	&	$	1.476	\pm	0.44	$	&	9	\\
61006	&	0.4377	&	$	966	\pm	322	$	&	$	0.24	\pm	0.01	$	&	$	2.06	\pm	0.15	$	&	$	0.983	\pm	0.15	$	&	3	\\
71227	&	0.381	&	$	1000	\pm	31	$	&	$	0.5	\pm	0.034	$	&	$	0.443	\pm	0.139	$	&	$	0.1	\pm	0.01	$	&	12	\\
101224A	&	0.72	&	$	393.88	\pm	161	$	&	$	0.12	\pm	0.013	$	&	$	0.824	\pm	0.125	$	&	$		--		$	&	10	\\
\enddata
\tablecomments{
Reference: (1)\citet{2007ApJ...671..656B} and references therein; (2)\citet{2005GCN..4394....1G}; (3)\citet{2006GCN..5710....1G}; (4)\citet{2007GCN..6638....1O}; (5)\citet{2013MNRAS.431.1398T}; (6)\citet{2010GCN.11470....1G}; (7)\citet{2016GCN.19843....1S}; (8)\citet{2016GCN.19570....1H}; (9)\citet{2013GCN.14771....1G}; (10) \citet{2010GCN.11489....1M}; (11)\citet{2010GCN.10912....1B}; (12)\citet{2007GCN..7155....1G}; (13)\citet{2007GCNR...74....2Z}; (14)\citet{2005GCN..3385....1B}.
（15）\citet{2019arXiv190205489W}  (16)\citet{2018ApJ...852....1Z} (17)\citet{2013MNRAS.430..163Q}.
}
\end{deluxetable*}
\clearpage

\end{document}